\documentclass[11pt,a4paper]{article}
\usepackage{graphicx}
\usepackage{latexsym}
\usepackage{subfigure}
\usepackage{amsmath}
\usepackage{enumerate}
\usepackage{amssymb}
\usepackage[mathscr]{eucal}
\usepackage{amsthm}
\usepackage{color}
\usepackage[small]{caption}
\usepackage[hypertexnames=false,colorlinks=true,linkcolor=blue,citecolor=blue,
            bookmarksopen=false,bookmarks=false,
            pdfstartview=XYZ,pdffitwindow=true,pdfcenterwindow=true]{hyperref}
\usepackage[a4paper,text={6.0in,9.0in},centering,
            includefoot,foot=0.5in]{geometry}
\usepackage[printfigures]{figcaps}
\usepackage{authblk}
\allowdisplaybreaks[1]

\newtheorem{thm}{Theorem}[section]

\theoremstyle{definition}

\newtheorem{model}[thm]{Model problem}

\theoremstyle{remark}

\renewcommand{\d}{\mathrm{d}}

\newcommand{\x}{\mathbf{x}}
\newcommand{\X}{\mathbf{X}}
\begin{document}

\figcapsoff

\title{Coarse-grained computation of traveling waves of lattice Boltzmann models with
Newton--Krylov solvers}

\author[1]{Giovanni Samaey}
\author[1]{Wim Vanroose}
\author[1]{Dirk Roose}
\author[2]{Ioannis G. Kevrekidis}
\affil[1]{\small{
Dept. of Computer Science, K.U. Leuven,
Celestijnenlaan 200A, 3001 Leuven, Belgium}}
\affil[2]{\small{
Dept. of Chemical Engineering and PACM,
Princeton University, Princeton, NJ08544}}

\date{\today}

\maketitle

\begin{abstract}
For many complex dynamical systems, a separation of scales prevails
between the (microscopic) level of description of the available model,
and the (macroscopic) level at which one would like to observe and
analyze the system.  For this type of problems, an ``equation-free''
framework has recently been proposed.  Using appropriately initialized
microscopic simulations, one can build a coarse-grained time-stepper to
approximate a time-stepper for the unavailable macroscopic model.  Here, we  
show how one can use this coarse-grained time-stepper to compute coarse-grained
traveling wave solutions of a lattice Boltzmann model.  In a moving frame,
emulated by performing a shift-back operation after the coarse-grained
time-step, the traveling wave appears as a steady state, which is
computed using an iterative method, such as Newton--GMRES.
To accelerate convergence of the GMRES procedure, a macroscopic
model-based preconditioner is used, which is derived from
the lattice Boltzmann model using a Chapman--Enskog expansion.
We illustrate the approach on a lattice Boltzmann model for the Fisher
equation and on a model for ionization waves.

\end{abstract}
\clearpage

\section{Introduction \label{sec:introduction}}

There is an established algorithmic infrastructure to study the long-term
dynamical features of systems of partial differential equations (PDEs), such as
steady states or periodic solutions.  When only a simulation code (a
time-stepper) is available, algorithms such as the recursive projection method
(RPM) \cite{SchroffKel93} and Newton--Picard \cite{KurtPHD, LustRooSpenChamp98}
can locate steady states, as well as their stability, and perform a
continuation for changing values of the parameters.  These methods 
project the Jacobian onto the eigenspace corresponding to the slowly decaying modes, 
which is typically low-dimensional.
In this subspace, a Newton iteration is performed; in the orthogonal complement, Picard iterations
(time-stepping) converge fast enough to the steady state.
 Alternatively, so-called Jacobian-free Newton--Krylov methods \cite{KnollKeyes04} solve
the linear system for each Newton correction by means of an iterative method,
such as GMRES, for which an appropriate preconditioner is crucial.  
Both standard preconditioning techniques, such as incomplete LU factorization (ILU)
and multigrid, as well as application-specific physics-based preconditioners have been proposed,
see \cite{KnollKeyes04} for an overview and references. If required, 
the stability can be computed as a post-processing step \cite{LeHSal01}.
As a common feature, all these methods only use selected
matrix-vector products with the system's Jacobian, which are estimated using
the time-stepper with several nearby initial conditions.

Unfortunately, a low-dimensional macroscopic PDE
is often not able to capture all detailed physical interactions accurately.  In such cases,
one needs to resort to a more microscopic description.  For instance,  
the dynamics of a system of colliding particles with interactions 
that depend sensitively on the relative particle velocities can, in general,
not be modeled by a reaction-diffusion equation for the particle density. 
One example, which forms the main motivation for the present paper, is the
impact ionization reaction, where each collision of an electron with a neutral
atom or molecule creates an additional electron when the relative velocity is
above a certain treshold.  Such a dynamical system should be modeled through a
phase space evolution law, e.g.~a Boltzmann equation.

Nevertheless, a clear separation in time-scales is often present in the microscopic
model: on fast time-scales, the microscopic variables equilibrate with respect
to a few macroscopic variables, while these macroscopic variables themselves
evolve on much slower time-scales.   When this is the case, a macroscopic model
should conceptually exist.  However, it might be difficult (or impossible) to 
derive a closed expression from the underlying microscopic model without introducing
assumptions that are hard to justify.  

For such models, there is an active current interest in the development of
so-called \emph{equation-free methods} to study the long-term behavior
\cite{KevrGearHymKevrRunTheo03}.  The key idea, which was first proposed in
\cite{TheoQianKevr00}, is to construct a \emph{coarse-grained time-stepper} for
the unavailable macroscopic equation as a three step procedure: (1) lifting,
i.e.\ the creation of appropriate initial conditions for the microscopic model,
conditioned upon the macroscopic state at time $t^*$; (2) simulation, using the
microscopic model, over the time interval $[t^*,t^*+\delta t]$; (3)
restriction, i.e.\ the extraction of the macroscopic state at time $t^*+\delta
t$.  The result is a coarse time-$\delta t$ map, which can be used to
estimate the matrix-vector products that are required in an RPM, Newton--Picard
or Newton--Krylov method.  

Based on RPM, coarse-grained bifurcation analysis has already been used in a
number of applications \cite{HumKevr03, SietGrahKevr03}, and also allows to
perform other system-level tasks, such as control and optimization
\cite{SietArmMakKevr03}.  In this paper, we will investigate the use of
Jacobian-free Newton--Krylov techniques on a model problem concerning
traveling wave solutions of lattice Boltzmann models.  Traveling waves
are solutions that move with constant speed without changing shape; in
a co-moving frame, they appear as steady state solutions.  We
construct a coarse-grained time-stepper in this co-moving frame by
performing a shift-back operation after each coarse-grained time-step,
and compute its fixed points using a Newton--GMRES procedure.  To
accelerate convergence of the GMRES procedure, we build a preconditioner based on an
approximate PDE model, which is derived from the lattice Boltzmann
model through a Chapman--Enskog expansion.  
As a consequence, the method described here could more appropriately be called
\emph{equation-assisted}, rather than equation-free.
We expect that the
techniques described here can be applied in other applications where particle
based methods are necessary to describe the dynamics.

This paper is organized as follows.  In section \ref{sec:2}, we briefly review
the basic properties of the coarse-grained time-stepper.  Subsequently, we outline
the model problems that will be used throughout the text in section \ref{sec:3}.
The non-linear system, of which the traveling waves are the solution, is constructed
in section \ref{sec:4}, and the preconditioned Newton--GMRES method is discussed in section
\ref{sec:5}.  Section \ref{sec:6} contains a detailed numerical study of the convergence
properties of the method.  Finally, we conclude in section \ref{sec:7}, which
contains a discussion of the computational complexity and some final remarks.

\section{Coarse-grained time-stepper\label{sec:2}}
We briefly review the coarse-grained time-stepper, as it was introduced by Kevrekidis
\emph{et al.} \cite{KevrGearHymKevrRunTheo03}.
To this end, we consider an abstract microscopic evolution law,
\begin{equation}\label{eq:micro_model}
\partial_t u(\x,t) = f(u(\x,t)),
\end{equation}
in which $u(\x,t)$ represents the microscopic state variables, $\x\in
D_m$ and $t$ are the microscopic independent variables, and
$\partial_t$ denotes the time derivative.  We assume that a macroscopic model,
denoted by
\begin{equation}\label{eq:macro_model}
\partial_t U(\X,t) = F(U(\X,t)),
\end{equation}
conceptually exists, but is unavailable in closed form. In equation (\ref{eq:macro_model}), 
$U(\X,t)$ represents the macroscopic
state variables, and $\X \in D_M$ and $t$ are the macroscopic
independent variables.

We introduce a time-stepper $s$ for the microscopic evolution law (\ref{eq:micro_model}),
\begin{equation}\label{eq:intro_micro_timestepper}
u(\x,t+\d t)= s(u(\x,t);\d t),
\end{equation}
where $\d t$ is the size of the microscopic time-step, and the aim is
to obtain a
coarse-grained time-stepper $\bar{S}$ for the variables $U(\X,t)$ as
\begin{equation}\label{eq:intro_macro_timestepper}
\bar{U}(\X,t+\delta t)=\bar{S}(\bar{U}(\X,t);\delta t),
\end{equation}
where $\delta t$ denotes the size of the coarse-grained time-step, and the
bars have been introduced to emphasize the fact that the time-stepper for the
macroscopic variables is only an \emph{approximation} of a time-stepper for
(\ref{eq:macro_model}), since this equation is not explicitly known.

To define a coarse-grained time-stepper
(\ref{eq:intro_macro_timestepper}), we need to introduce two operators
that make the transition between microscopic and macroscopic variables.  We
define a \emph{lifting operator},
\begin{equation}\label{eq:intro_lifting}
\mu: U(\X,t) \mapsto u(\x,t)=\mu(U(\X,t)),
\end{equation}
which maps macroscopic to microscopic variables, and its complement, the
\emph{restriction operator},
\begin{equation}\label{eq:intro_restriction}
\mathcal{M}: u(\x,t) \mapsto U(\X,t)=\mathcal{M}(u(\x,t)).
\end{equation}
The restriction operator can often be determined as soon as the
macroscopic variables are known.  For instance, when the microscopic
model consists of an evolving ensemble of many particles, the
restriction typically consists of the computation of the first few
low order moments of the distribution (density, momentum, energy),
which are considered as the appropriate macroscopic variables
$U(\X,t)$, in terms of which a closed macroscopic equation can be
written. The assumption that a macroscopic equation closes at the level of these
low order moments, implies that the higher order moments become
functionals of the low order moments on time-scales which are fast
compared to the overall system evolution (\emph{slaving}).

The construction of the lifting operator is usually more involved.
Again taking the example of a particle model, we need to define a
mapping from a few low order moments to initial conditions 
for each of the particles.  We know
that the higher order moments of the resulting particle distribution
should be functionals of the low order moments, but unfortunately,
these functionals are unknown (since the macroscopic evolution law is
also unknown).  Several approaches have been suggested to address this
problem.  One could for instance initialize the higher order moments
randomly.  This introduces a
\emph{lifting error}, and one then relies on the separation of time-scales to
ensure that the higher order moments relax quickly to a functional of the
low order ones (\emph{healing}) \cite{GearKevrTheo02, MakMarPanKevr02,
SietGrahKevr03} (see also \cite{Ciccotti, Torrie}).   We note that, in some cases, this approach 
produces inaccurate results \cite{PvLLustKevr05}.  In fact, to initialize the
higher order moments correctly, one should perform a simulation of the
microscopic system subject to the constraint that the low order moments
are kept fixed.  How this can be done using only a time-stepper for the
original microscopic system, is explained and analyzed in
\cite{GearKapKevrZag05, Constraint03, PvLWimRoo05}.  We will briefly discuss the
lifting step for our model problems in section \ref{sec:cg_ts}.

Given an initial condition for the macroscopic variables $U(\X,t^*)$ at some
time $t^*$, we can construct the time-stepper
(\ref{eq:intro_macro_timestepper}) in the following way:
\begin{enumerate}
\item \textbf{Lifting.} Using the
lifting operator (\ref{eq:intro_lifting}), create appropriate
initial conditions $u(\x,t^*)$ for the microscopic time-stepper
(\ref{eq:intro_micro_timestepper}), consistent with the
macroscopic variables.
\item \textbf{Simulation.} Use the microscopic time-stepper
(\ref{eq:intro_micro_timestepper}) to compute the microscopic state $u(\x,t)$
for $t \in [t^*,t^*+\delta t]$.
\item \textbf{Restriction.} Obtain the macroscopic state $U(\X,t^*+\delta t)$
from the microscopic state $u(\x,t^*+\delta t)$ using the restriction operator
(\ref{eq:intro_restriction}).
\end{enumerate}
Assuming $\delta t = k \d t$, this can be written as
\begin{equation}\label{eq:intro_coarse_timestepper}
\bar{U}(\X,t+\delta t)=\bar{S}(\bar{U}(\X,t),\delta t)=\mathcal{M}(s^k(\mu(\bar{U}(\X,t)),\d t)),
\end{equation}
where we have represented the $k$ microscopic time-steps by a superscript on
$s$.  If the microscopic model is stochastic, one may need to perform multiple
replica simulations, using an ensemble of microscopic initial conditions, to obtain 
an accurate result with a sufficiently low variance.

\section{Model problems\label{sec:3}}
We now briefly discuss the origins of the lattice Boltzmann method and the relation
with the Boltzmann equation; we also introduce our model problems.

\subsection{Microscopic and macroscopic model}

We consider systems that, on a molecular level, consist of particles
whose position and velocity are governed by two processes: free flight
and collisions.  We introduce the probability $f(x,v,t)\d x\d v$,
which represents the fraction of particles with position and velocity
in the infinitesimal domain $[x,x+\d x]\times [v,v+\d v]$ at time $t$.
The evolution of $f$ is governed by a so-called kinetic equation,
\begin{equation}\label{eq:boltz}
\partial_t f(x,v,t) + v \partial_x f(x,v,t) + F \partial_v f(x,v,t) = Q(f,f),
\end{equation}
where $Q(f,f)$ is the \emph{collision integral}, and $F$ is an
external force term.  The first equation of this type was the 
\emph{Boltzmann equation} for moderately rarefied gas flows with
$F\equiv 0$ \cite{boltz}.  Of course, when multiple species are
present, (\ref{eq:boltz}) becomes a system of equations.  Throughout this paper,
we confine ourselves to problems in one space dimension.

Usually, one introduces a \emph{kinetic model} for $Q(f,f)$ to
simplify equation (\ref{eq:boltz}) \cite{Gorban}.  A standard choice
is the non-linear Bhatnagar--Gross--Krook model (BGK) \cite{bgk},
\begin{equation}\label{eq:bgk}
\partial_t f(x,v,t) + v \partial_x f(x,v,t) + F \partial_v f(x,v,t) = -\frac{f(x,v,t)-f^{eq}(x,v,t)}{\tau}.
\end{equation}
Here, the collisions are interpreted as a relaxation to local
equilibrium distribution, with a characteristic relaxation time $\tau$.  The choice
of the equilibrium distribution $f^{eq}(x,v,t)$ (and therefore of the
collision integral $Q(f,f)$) determines the physics of the system.

Although the kinetic equation (\ref{eq:boltz}) and its BGK-approximation are able to describe the evolution
of a wide range of physical systems, a numerical simulation quickly becomes intractable because
of the high dimensionality.  However, one can often obtain an approximate macroscopic description.  We define the \emph{moments}
of the distribution function as
\begin{align}
\bar{\rho}^{(i)}(x,t)&=m\int_{-\infty}^{\infty}v^i f(x,v,t)\d v, & i = 0, 1, 2, \ldots,
\end{align}
where $m$ is the particle mass.
The lowest order moments correspond to the density ($i=0$), momentum ($i=1$) and energy ($i=2$),
which we write as
\begin{equation}\label{eq:mom}
\rho(x,t)=\bar{\rho}^{(0)}(x,t), \qquad \phi(x,t)=\bar{\rho}^{(1)}(x,t), \qquad
\xi(x,t)=\bar{\rho}^{(2)}(x,t)/2.
\end{equation}
Using the assumption that the deviation from local equilibrium is sufficiently small, one performs an asymptotic
expansion (the Chapman--Enskog expansion, \cite{Liboff03}) to obtain a closed description for the evolution
of a number of low order moments.  If we define the macroscopic variables
as $U(x,t)=\left\{\bar{\rho}^{(i)}(x,t)\right\}_{i=0}^{M}$, we obtain an approximate PDE of the form
\begin{equation}\label{eq:mom_eq}
\partial_t U(x,t)=F\left(U(x,t),\partial_x U(x,t),\ldots,\partial_x^d U(x,t)\right),
\end{equation}
which depends on the first $d$ spatial derivatives.
Equation (\ref{eq:mom_eq}) is a good approximation of the system dynamics 
when there is a fast decay of the higher order terms in the Chapman--Enskog expansion.

Consider as an example the advection-diffusion equation,
\begin{equation}\label{eq:advdiff}
\partial_t \rho(x,t) + c \partial_x \rho(x,t) = \partial_x (D \partial_x \rho(x,t) ),
\end{equation}
with transport coefficient $c$ and diffusion coefficient $D$.
This equation can be derived from equation (\ref{eq:bgk}) with $F\equiv 0$
by defining the equilibrium distribution as
\begin{displaymath}
f^{eq}(x,v,t) = \rho(x,t)\sqrt{\frac{\lambda}{\pi}}\exp(-\lambda(v-c)^2),
\end{displaymath}
with $\lambda = m / 2kT$, where $T$ is temperature and $k$ is the Boltzmann constant,
and adding the conservation constraint
\begin{displaymath}
\int_{-\infty}^{\infty}\left(f(x,v,t)-g(x,v,t)\right)\d v = 0.
\end{displaymath}
A straightforward derivation reveals that the macroscopic behaviour
of (\ref{eq:bgk}) is described by equation (\ref{eq:advdiff}) when
we choose $\tau = 2 D \lambda$ \cite{XuVKI98}.  Different equilibrium distributions can be
used to obtain the Burgers' equation, the Euler equations, the
Navier--Stokes equations, etc.  One can also add chemical reactions in
the collision integral \cite{Gorban}.

In this paper, we will use the \emph{lattice Boltzmann method} (LBM) \cite{ChopDupMasLut02, Succi2001},
which can be viewed as a special discretization of
the Boltzmann equation \cite{HeLuo97}.  In an LBM method, the distribution function $f(x,v,t)$ is discretized on a
space-time lattice with grid spacing $\d x$ in space and $\d t$ in time.  Only a discrete
number of velocities are considered, which correspond to a movement over an integer number of
lattice points during one time-step,
\begin{displaymath}
v_i = c_i \frac{\d x}{\d t}, \qquad c_i = -q, -q + 1, \ldots, q-1, q.
\end{displaymath}
For ease of exposition, we restrict ourselves to the case $q=1$, which gives only three speeds (the so-called D1Q3 model).

For reaction-diffusion systems, we can write the evolution law for $f_i(x,t)=f(x,v_i,t)$ as
\begin{equation}\label{eq:lbm}
f_i(x+c_i \d x, t +\d t) = (1-\omega)f_i(x,t)-\omega f_i^{eq}(x,t)+R_i(x,t), \qquad i = -1,0,1.
\end{equation}
The right-hand side approximates the collision operator, and is composed of a reaction term
$R_i(x,t)$ and a BGK relaxation to the local equilibrium.
After collision, the post-collision values are propagated to a neighboring lattice site, which corresponds to free flight.

For the lattice Boltzmann discretization, we can define the (non-dimensional) moments of the distribution function as
\begin{equation}\label{eq:lbm_mom}
\rho(x,t)=\sum_{i=-1}^{1}f_i(x,t),\qquad \phi(x,t)=\sum_{i=-1}^{1}c_i f_i(x,t),
\qquad \xi(x,t)=\frac{1}{2}\sum_{i=-1}^{1}c_i^2 f_i(x,t).
\end{equation}
It can easily be verified \cite{DawChenDoo93, QianOrs95} that, under suitable smoothness assumptions, the system is well approximated
by a macroscopic reaction-diffusion equation
\begin{equation}\label{eq:rd}
\partial_t \rho(x,t) = \partial_x(D\partial_x \rho(x,t))+r(\rho(x,t)),
\end{equation}
which can again be derived from the LBM equation using a Chapman--Enskog expansion
\cite{ChopDupMasLut02, PvLWimRoo06} using
\begin{equation}\label{eq:equiv}
\omega = \frac{2}{1+3D\d t/\d x^2}, \qquad R_i(x,t)=\frac{\d
t}{3}r\left(\rho(x,t)\right),\qquad f_i^{eq}(x,t)=\rho(x,t)/3.
\end{equation}

To view these models in the equation-free context, we define
\begin{displaymath}
u(\x,t)=u(x,t)=\left\{f_i(x,t)\right\}_{i=-1}^{1},\qquad  U(\X,t)=U(x,t)=\rho(x,t).
\end{displaymath}

\subsection{Model problems and traveling waves}

In this paper, we propose a numerical method to compute
coarse-grained traveling wave solutions of lattice Boltzmann
models of the form (\ref{eq:lbm}). Traveling waves are solutions of the form
\begin{equation}
U(x,t) = U(x - c t) = U(\zeta), \qquad \lim_{\zeta \to \pm \infty} U(\zeta)=U_{\infty}^{\pm},
\end{equation}
where $\zeta = x - c t$.  They appear as steady states in a moving frame $(\zeta,t)$.

In this work,
we consider traveling fronts, for which $U_{\infty}^-\neq U_{\infty}^+$.  For these solutions,
 the evolution of the density is not sufficiently well described by the approximate reaction-diffusion
PDE (\ref{eq:rd}), due to a lack of smoothness of the density in the reaction front.  
Nevertheless, we also observe in this case that the higher order moments (momentum and energy)
become (more complicated) functionals of the density on fast time-scales. As a consequence,
a PDE for the density should still exist.

We consider two model problems: a model
which is derived from the Fisher PDE,
and a lattice Boltzmann model for ionization.  Both model problems
exhibit traveling fronts with arbitrary speed $c \ge c^*$, where $c^*$ is called the
critical speed.

\begin{model}[Fisher model]\label{mod:fisher}
We consider the reaction-diffusion lattice Boltzmann model,
\begin{equation}\label{eq:lbm_fish} f_i(x+c_i\d x, t +\d t) =
(1-\omega)f_i(x,t)-\omega f_i^{eq}(x,t)+R_i(x,t),
\end{equation}
where $i = -1,0,1$, with
\begin{equation}\label{eq:lbm_fish_reac}
R_i(x,t)=\frac{\d t}{3}\rho(x,t)(1-\rho(x,t)).
\end{equation}
Following the reasoning of the previous section, we define the approximate macroscopic PDE,
\begin{equation}\label{eq:fish}
\partial_t \rho(x,t) = \partial_x(D\partial_x\rho(x,t))+\rho(x,t)(1-\rho(x,t)),
\end{equation}
which was originally proposed by Fisher as a model for the spread of
advantageous genes \cite{fish}.  This equation appears in a range of physical
and biological applications exhibiting waves, see e.g. \cite{murray}, and supports traveling fronts of the form
$\rho(x,t)=\rho(\zeta)$, with
\begin{displaymath}
\lim_{\zeta \to -\infty}\rho(\zeta)=1, \qquad \lim_{\zeta \to \infty}\rho(\zeta)=0.
\end{displaymath}

This model problem can be put in the equation-free framework by
identifying
\begin{displaymath}
u(\x,t)=u(x,t)=\left\{f_i(x,t)\right\}_{i=-1}^{1},\qquad
U(\X,t)=U(x,t)=\rho(x,t).
\end{displaymath}


The Fisher equation is arguably the simplest model problem that exhibits traveling waves
of arbitrary speed.  It is known for the corresponding PDE that traveling waves
exist with speed $c\ge c^*=2\sqrt{D}$.  Numerically, traveling waves exists for
all $c$.  However, the critical speed $c^*$ is the lowest speed for which the
traveling waves are uniformly positive.  The coarse-grained traveling wave
with minimal speed $c^*=2\sqrt{D}$ is shown in figure \ref{fig:wave_fish}.
\begin{figure}
\begin{center} \includegraphics[scale=0.8]{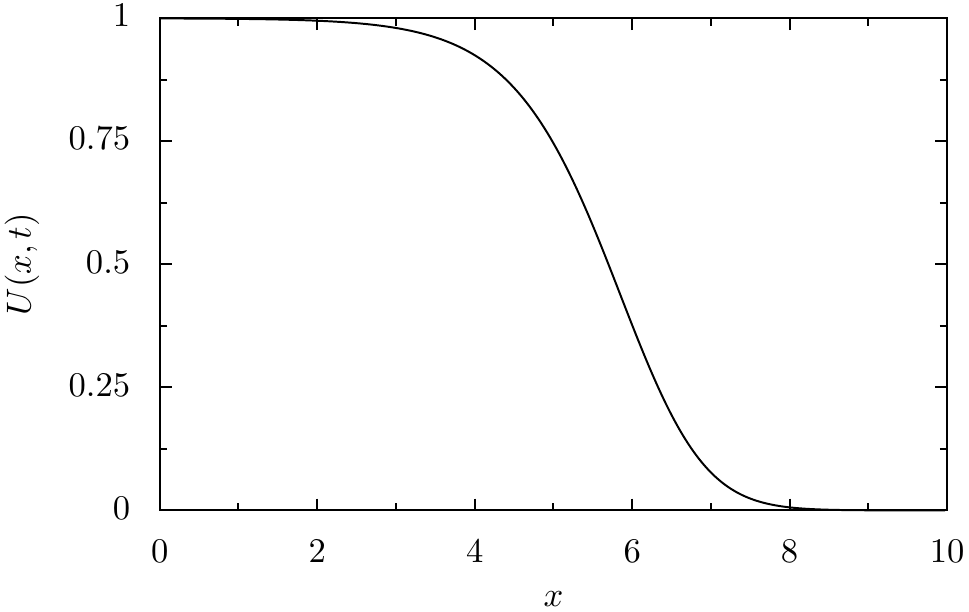} \end{center}
\caption{\label{fig:wave_fish} Coarse-grained traveling wave solution of the Fisher
lattice Boltzmann model (\ref{eq:lbm_fish})-(\ref{eq:lbm_fish_reac}).}
\end{figure}

\end{model}

\begin{model}[Planar streamer fronts]\label{mod:ionization}
Streamers are sharp, non-linear waves of electrons that propagate
through gases in the presence of strong electric fields. The strong
field accelerates the electrons that cause ionization reactions during
the collisions with the neutral gas particles. This ionization reaction
creates additional electrons that are, again, accelerated. This
results in an avalanche at the tip of the wave front.

We consider a lattice Boltzmann model for the distributions $f_i(x,t)$ of electrons,
which is coupled to a PDE that governs the evolution of the electric field
through the electron density $\rho(x,t)=\sum_i f_i(x,t)$. 
The coupled equations are
\begin{equation}\label{eq:lbm_ebert}
\left\{\begin{array}{rcl}
f_i(x+c_i \d x,t+\d t) &=& (1-\omega)f_i(x,t) - \omega f_i^{eq} (x,t) - E(x,t) \sum_j V_{ij} f_j(x,t)  +  R_i(x,t),  \\
\partial_t E(x,t)  &=& - \rho(x,t) E(x,t) - D \partial_x \rho(x,t)
       \end{array}
\right.
\end{equation}
in which the electric field $E(x,t)$ appears as an external force. In the
original Boltzmann equation \eqref{eq:boltz}, this force appears as
$E(x,t)\partial_v f(x,v,t)$;
in the lattice Boltzmann equation, this external force is discretized as $E(x,t)\sum_j V_{ij} f_j$,
as proposed by Luo \cite{luo}.

In general, the reaction terms $R_i(x,t)$ should be modeled on a microscopic level.
However, due to reasons of computational complexity,
most analysis of streamers is based on the Townsend approximation
of the microscopic ionization reactions that appear at the tip of the front \cite{Ebert}.
The reaction terms are then given by $R_i(x,t)=\d t\rho(x,t) g(E(x,t))/3$, with
$g(E) = |E| \exp(-1/|E|)$.

In this paper, we will also use this Townsend approximation to analyze the performance 
of our numerical method.  In a forthcoming publication, the method presented here will be used to 
analyze the traveling waves of a five-speed lattice Boltzmann model, which is based on a more
realistic set of microscopic interactions \cite{WimSamPvLRoo06}.

Again, we can derive an approximate PDE model,
\begin{equation}\label{eq:ebert_pde}
\left\{
\begin{aligned}
\partial_t\rho(x,t) &=  \partial_x (\rho(x,t) E(x,t)) + D \partial_x^2 \rho(x,t) + \rho(x,t) g(E(x,t)),\\
\partial_t E(x,t) &= - \rho(x,t) E(x,t) - D \partial_x \rho(x,t),
\end{aligned}
\right.
\end{equation}
This coupled equation exhibits traveling wave
solutions with arbitrary speed $c \ge c^* =
\lim_{x \rightarrow \infty} |E(x)| + 2\sqrt{D g(E(x))}$, see figure \ref{fig:wave_ebert}.

\begin{figure}
\begin{center}
\includegraphics[scale=1]{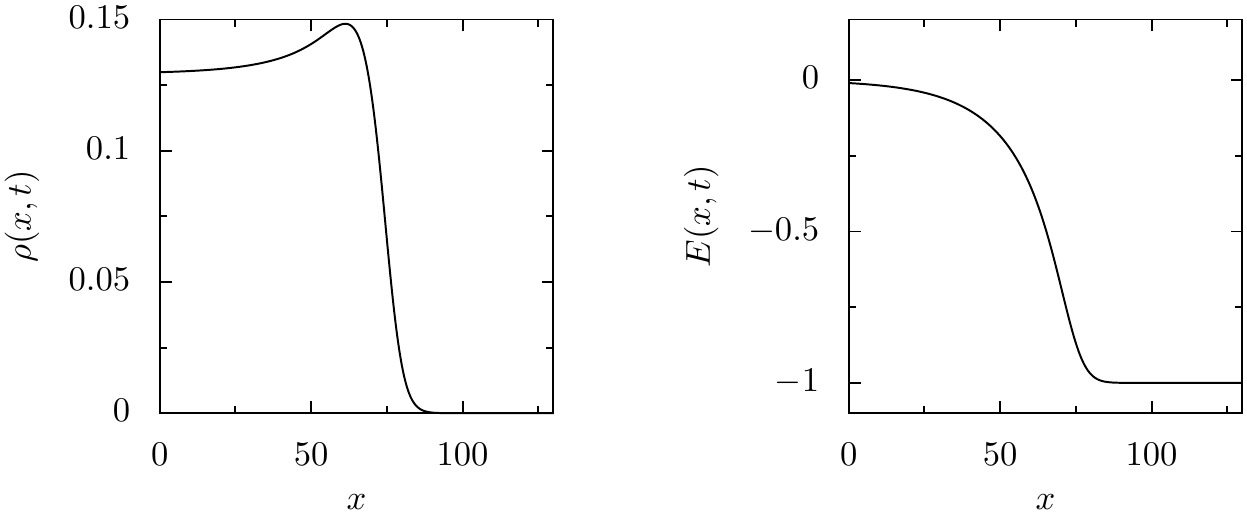}
\end{center}
\caption{\label{fig:wave_ebert}
 Coarse-grained traveling wave solution of the lattice Boltzmann model
 for planar streamer fronts (\ref{eq:lbm_ebert}). We show both the density $\rho(x,t)$ (left)
and the electrical field $E(x,t)$ (right).}
\end{figure}

This model problem can be put in the equation-free framework by
considering
\begin{displaymath}
u(\x,t)=u(x,t)=\left\{\{f_i(x,t)\}_{i=-1}^{1},
E(x,t)\right\},\qquad U(\X,t)=U(x,t)=\left\{
\rho(x,t),E(x,t)\right\}.
\end{displaymath}

\end{model}

\section{The fixed point problem \label{sec:4}}

To define the fixed point problem for the traveling wave, we 
first construct a coarse-grained time-stepper for
the lattice Boltzmann model of the form
\begin{equation}
\bar{U}(x,t+\delta t)=\bar{S}(\bar{U}(x,t),\delta t),
\end{equation}
where the coarse-grained time-step $\delta t = k \d t$,
see equation (\ref{eq:intro_macro_timestepper}).  
In section \ref{sec:cg_ts}, we describe the details of the lifting operator.

Traveling waves appear as a one-parameter family of solutions:
together with $\bar{U}^*(\zeta)$, any translate
$\bar{U}^*(\zeta+\varphi)$, for an arbitrary but fixed $\varphi$,
is also a traveling wave.  We therefore add a phase condition to
remove this indeterminacy.  The specific implementation of the shift-back
operator and the phase condition are discussed in section \ref{sec:phase}.

Note that we consider model problems for which a traveling wave
exists for arbitrary $c\ge c^*$. It is of physical interest to study
how the critical speed $c^*$ depends on the system parameters.
However, in this paper we confine ourselves to the computation of
a traveling wave for a fixed speed $c$.  A detailed analysis of the
physical properties of the ionization model can be found in \cite{WimSamPvLRoo06}.

\subsection{Lifting for lattice Boltzmann models\label{sec:cg_ts}}

As already outlined in section \ref{sec:2}, a coarse-grained time-stepper
is constructed as a lift-simulate-restrict procedure.  While the
restriction step is well defined by equation (\ref{eq:lbm_mom}), the lifting step
can be defined in multiple ways.  We discuss three possibilities.

\paragraph{Weighted lifting.}  A first possibility is to simply
initialize the distributions as \begin{equation}
f_i(x,t)=w_i \rho(x,t), \qquad \sum_i w_i =1,
\end{equation}
where $w_i = 1/3$ is an obvious choice, since this corresponds to
the local diffusive equilibrium as defined in equation
(\ref{eq:equiv}). As a consequence, the higher order moments are
initialized as
\begin{displaymath}
\phi(x,t)=0, \qquad \xi(x,t)=\rho(x,t)/3.
\end{displaymath}
This very rough approximation of the higher order moments
introduces a \emph{lifting error}, and one relies on the
separation of time-scales between low order and high order moments
to \emph{heal} this error quickly \cite{TheoQianKevr00}.  In
\cite{PvLLustKevr05}, it is shown how this initialization can
produce undesired artifacts for lattice Boltzmann models when $\delta t$ is small.

\paragraph{Slaving relations.}  From the assumption that a macroscopic model for $\rho(x,t)$
exists, it follows that the higher order moments $\phi(x,t)$
and $\xi(x,t)$ can be written as functionals of
$\rho(x,t)$.  For lattice Boltzmann models of the form (\ref{eq:lbm}) with (\ref{eq:equiv}),
 these so-called \emph{slaving relations} can be
written down analytically as an asymptotic expansion in
$1/\omega$. Up to third order, we have \cite{PvLWimRoo05},
\begin{equation}\label{eq:slaving}
\begin{aligned}
\phi(x,t)&=-\frac{2\d x}{3\omega}\partial_x\rho(x,t)+
\frac{\d x\d t}{\omega^2}\left(\frac{\omega}{\omega -
2}+\frac{1}{3}\right)
\left(\partial_x r\left(\rho(x,t)\right)-\partial^2_{x t} \rho(x,t)\right),\\
\xi(x,t)&=\frac{1}{3}\rho(x,t)-\frac{\d t}{6 \omega}
\left(r\left(\rho(x,t)\right)-\partial_t
\rho(x,t)\right).
\end{aligned}
\end{equation}
These expansions can alternatively be written down in terms of
$\rho(x,t)$ and its spatial derivatives only, by making use
of (\ref{eq:rd}). The weighted lifting scheme coincides with
the zeroth order term of the slaving relations.

\paragraph{Constrained runs scheme.} Although ideally one would like to use the slaving relations to
initialize the lattice Boltzmann time-stepper, this is not always
possible.  The analytical derivation is cumbersome, and might even
be impossible when the lattice Boltzmann model is coupled to a
PDE, as in model problem \ref{mod:ionization}.  Moreover, the
number of higher order moments increases when a more detailed
discretization of the velocity space is used.
 Finally, the analytic derivation of the slaving relations
is only tractable if only a small number of terms of the
asymptotic expansion are needed.  However, the macroscopic equation
becomes invalid precisely when a
large number of terms is needed.

\begin{figure}
\centering \setlength{\unitlength}{1cm}
\begin{picture}(12,6)
\thicklines \put(2.7,0.5){\vector(1,0){7.3}}
\put(2.7,0.5){\vector(0,1){5.5}} \qbezier[60](3,2.1)(4.5,3)(9,1)
\qbezier(3,1.8)(4.5,2.7)(9,0.7) \put(8,1.5){\centering
approximate} \put(7.6,0.7){\centering slow} \thinlines
\put(0.5,5){\centering $\{ \phi^{(0)} , \xi^{(0)} \}$}
\put(0.5,2.3){\centering $\{ \phi^* , \xi^* \}$}
\put(-0.1,1.8){\centering $\{ \phi(\rho^{(0)}) ,
\xi(\rho^{(0)}) \}$} \put(2.6,5){\line(1,0){0.2}}
\put(2.6,2.3){\line(1,0){0.2}} \put(2.6,2){\line(1,0){0.2}}
\put(4.9,0){\centering $\rho^{(0)}$}
\put(5,0.3){\line(0,1){4.7}} \put(5.9,0){\centering
$\rho^*$} \put(6,0.4){\line(0,1){0.2}} \thicklines
\put(4.6,4.4){\vector(0,-1){1.2}} \put(5,5){\circle*{0.2}}
\put(5,2.3){\circle*{0.2}} \put(5,2){\circle*{0.2}}
\qbezier(5,5)(5,4.3)(7.2,3.8) \put(7.2,3.8){\vector(-1,0){2.2}}
\qbezier(5,3.8)(5,3.4)(6.6,3) \put(6.6,3){\vector(-1,0){1.6}}
\qbezier(5,3)(5,2.7)(6.2,2.5) \put(6.2,2.5){\vector(-1,0){1.2}}
\qbezier(5,2.5)(5,2.4)(6,2.3) \put(6,2.3){\vector(-1,0){1}}
\end{picture}
\caption{Sketch of the evolution of the constrained runs scheme.
We plot the higher order moments $\phi$ and $\xi$ are
plotted as a function of the density $\rho$.  After each
lattice Boltzmann step, $\rho$ is reset to the given
$\rho^{(0)}$.  The successive iterates converge to
$\{\phi^*, \xi^*\}$, which is an approximation to the
slaved state $\{\phi(\rho^{(0)}),
\xi(\rho^{(0)})\}$ as given by equation
(\ref{eq:slaving}). Figure reproduced from \cite{PvLWimRoo05}. \label{fig:constrained}}
\end{figure}
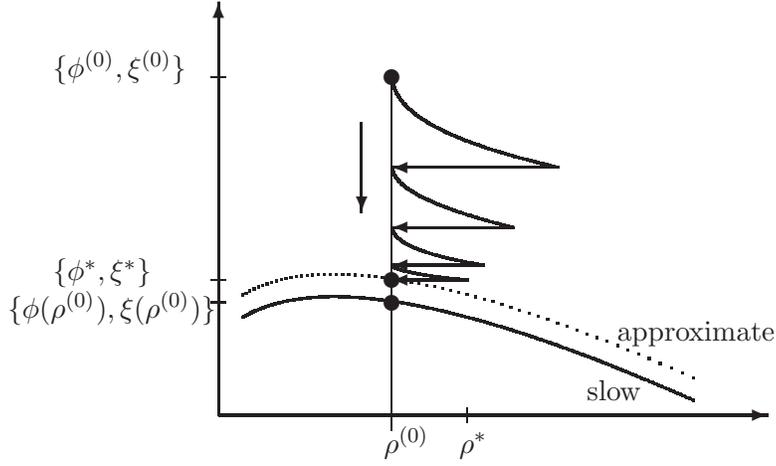

As an alternative, one can use the constrained runs scheme \cite{GearKapKevrZag05,
Constraint03} to approximate the full microscopic state corresponding to a
set of predescribed macroscopic variables.  The idea is to perform a
number of lattice Boltzmann time-steps using equation (\ref{eq:lbm_fish}),
where after each time-step the density is reset to the initial
density. During this iteration the microscopic variables converge
towards their relaxed values; correspondingly, the higher order moments have then approximately reached the
slaved state.
As such, the constrained runs scheme is a fixed point iteration for the higher
order moments of the microscopic model. 
We reproduce the schematic
representation that was given in \cite{PvLWimRoo05}
(figure
\ref{fig:constrained}).

The constrained runs scheme can readily be applied to any lattice
Boltzmann model, and defines a lifting operator that initializes the
microscopic variables very close to their relaxed values. We refer to
\cite{PvLWimRoo05} for a detailed convergence and error analysis
for reaction-diffusion systems.

\subsection{Shift-back procedure and phase condition\label{sec:phase}}

Next, we discuss our specific implementation of the shift-back
operator $\sigma_{\varphi}$.  By noting that a shift-back over a distance
$\varphi$ is equivalent with a time evolution of the transport equation
\begin{displaymath}
\partial_t U(x,t) - \partial_x U(x,t) = 0,
\end{displaymath}
over a time $\varphi$, we obtain
\begin{equation}\label{eq:shift}
\sigma_{\varphi}: U(x,t) \mapsto
\sigma_{\varphi}(U(x,t))=U(x,t) + \varphi\partial_x U(x,t),
\end{equation}
which is a valid approximation for  shift-backs over short distances.

At this point, we have constructed a non-linear system
\begin{equation}\label{eq:sing}
\bar{U}(\zeta) - \sigma_{c\delta t}\left(\bar{S}(\bar{U}(\zeta),\delta
t)\right)= 0,
\end{equation}
Unless we add a phase condition, this system is singular.  When defining
 the Jacobian of the coarse-grained time-stepper as
\begin{displaymath}
\bar{L}(\bar{U}(\zeta),\delta t) = \frac{\partial \left[\sigma_{c\delta
t}\left(\bar{S}(\bar{U}(\zeta),\delta t)\right)\right]}{\partial
\bar{U}(\zeta)},
\end{displaymath}
this singularity appears as a zero eigenvalue of
$I-\bar{L}(\bar{U}^*(\zeta),\delta t)$, with eigenfunction $\d
\bar{U}^*(\zeta)/\d \zeta$.

To compensate for the extra phase condition, we
add a regularization parameter $\alpha$ as an additional unknown, similar to what is done in
\cite{Schmidt} for Hamiltonian systems.
The resulting non-linear system is
\begin{equation}\label{eq:reg}
G(\bar{U},\alpha) = \left\{\begin{array}{lcr}
\bar{U} - \sigma_{c\delta t}\left(\bar{S}(\bar{U},\delta
t)\right) +\alpha \d_{\zeta} \bar{U} &=& 0,\\
p(\bar{U})&=&0,
\end{array}\right.
\end{equation}
where $\bar{U}$ denotes the space discretization of $\bar{U}(\zeta)$
and $p(\bar{U})$ is the phase condition defined below.  This problem
is well-posed for the unknowns $(\bar{U},\alpha)$ with a locally
unique solution $(\bar{U}^*,\alpha^*=0)$.

Here, we use the phase condition
\begin{displaymath}
p(\bar{U}) = \int_{\zeta_0}^{\zeta_{N-1}}\bar{U}(\zeta)\d_{\zeta}
\bar{U}_{ref}(\zeta)\d \zeta,
\end{displaymath}
which minimizes phase shift with respect to the reference solution
$\bar{U}_{ref}(\zeta)$, as is done in publicly available software for 
bifurcation analysis, such as AUTO (for ordinary differential equations)
\cite{phase,auto1,auto2} or DDE-BIFTOOL \cite{ddebif} (for delay differential equations).

\section{Preconditioned Newton--GMRES\label{sec:5}}
We solve the non-linear equation (\ref{eq:reg}) for the traveling wave solution.  First, we discretize $\bar{U}(\zeta)$ on the lattice Boltzmann grid
$[x_0,x_1=x_0+\d x,\ldots, x_{N-1}]$, and provide
the time-stepper with homogeneous Neumann boundary conditions.  The spatial
derivative in (\ref{eq:shift}) is discretized using central differences.

We solve the resulting non-linear system (\ref{eq:reg}) with a
Newton--Raphson procedure
\begin{equation}\label{eq:newt}
\left\{ \begin{array}{lcr}
\bar{U}^{(k+1)} &=& \bar{U}^{(k)}+\d\bar{U}^{(k)},\\
\alpha^{(k+1)} &=& \alpha^{(k)}+\d\alpha^{(k)},
\end{array}\right.
\end{equation}
where $\d\bar{U}^{(k)}$ and $\d \alpha^{(k)}$ are the corrections calculated
each iteration by solving linear systems of the form
\begin{equation}\label{eq:linear}
 A(\bar{U}^{(k)},\delta t) \begin{bmatrix}
\d \bar{U}^{(k)}\\
\d \alpha^{(k)}
\end{bmatrix}= \begin{bmatrix}
I-\bar{L}(\bar{U}^{(k)},\delta t) & \d_{\zeta} \bar{U}^{(k)}\\
\d_{\bar{U}} p\left(\bar{U}^{(k)}\right) & 0
\end{bmatrix}
\begin{bmatrix}
\d \bar{U}^{(k)}\\
\d \alpha^{(k)}
\end{bmatrix}
= -G(\bar{U}^{(k)},\alpha^{(k)}),
\end{equation}
where $A(\bar{U}^{k},\delta t)$ denotes the linearization of
$G(\bar{U},\alpha)$ around the point
$(\bar{U}^{(k)},\alpha^{(k)})$. The right hand side is the residual
at the same point.

We do not have an explicit formula for
$\bar{L}(\bar{U}^{(k)},\delta t)$, since it involves the Jacobian of
the coarse-grained time-stepper with shift-back.  However, we can
estimate matrix-vector products as
\begin{equation}\label{eq:matvec}
\left(I-\bar{L}(\bar{U},\delta t)\right)v\approx v -
\frac{\sigma_{c\delta t}\left(\bar{S}(\bar{U}+\epsilon v,\delta
t)\right) -\sigma_{c\delta t}\left(\bar{S}(\bar{U},\delta
t)\right)}{\epsilon}.
\end{equation}
Therefore, we solve the linear system (\ref{eq:linear}) using a Krylov
method, such as GMRES \cite{gmres}.

The convergence rate of GMRES depends sensitively  on the spectral
properties of the system matrix in equation (\ref{eq:linear}). 
For GMRES to converge rapidly, the eigenvalues should be clustered, e.g.\ 
around one \cite{saad}. 
It can be checked that the bordering row and column transform the
zero eigenvalue of the Jacobian of (\ref{eq:sing}) into $\pm\sqrt{\,
\d_{\bar{U}}p \cdot \d_{\zeta} \bar{U}}$, while leaving the 
other eigenvalues unaltered \cite{KurtPrive}.  The eigenvalues
$\mu_k$ of $1-\bar{L}(\bar{U},\delta t)$ are of the form
\begin{displaymath}
\mu_k = 1-\exp(\lambda_k \delta t),
\end{displaymath}
where $\lambda_k\approx O(-k^2)$ for $k\gg 1$.  For $k$ small,
$\lambda_k \approx 0$. 
When the system possesses a low-dimensional inertial manifold, one can choose
$\delta t$ large, such that the spectrum becomes a compact perturbation of the
unit matrix \cite{KelKevrQiao04}.  In this case, GMRES is known to converge
very quickly; this, however, at the cost of a long simulation time for each
matrix-vector product. If $\delta t$ is chosen to be small, the eigenvalues
$\mu_k$ range from $0$ to $1$.   In that case, preconditioning will be
necessary.

We define a preconditioning matrix $M(\bar{U},\delta t)$, and
replace the linear system (\ref{eq:linear}) with
\begin{equation}\label{eq:precond}
 M(\bar{U}^{(k)},\delta t)^{-1}
 A(\bar{U}^{(k)},\delta t)
\begin{bmatrix}
\d \bar{U}^{(k)}\\
\d \alpha^{(k)}
\end{bmatrix}= -M(\bar{U}^{(k)},\delta t)^{-1} G(\bar{U}^{(k)},\alpha^{(k)}).
\end{equation}
Ideally, $M(\bar{U},\delta t)$ is both a good approximation of the
system matrix, as well as a matrix for which an efficient (direct)
solver is available.  We will use a time-stepper for the
approximate macroscopic equation (\ref{eq:mom_eq}) to construct
$M(\bar{U},\delta t)$. Consider the macroscopic equation
(\ref{eq:mom_eq}) in the moving frame $(\zeta,t)$,
\begin{align}\label{eq:moving}
\partial_t U(\zeta,t) &= F\left(U(\zeta,t),
\partial_{\zeta}U(\zeta,t),\ldots,\partial_{\zeta}^d U(\zeta,t)\right)
+ c \partial_\zeta U(\zeta, t),\\
& = \tilde{F}\left(U(\zeta,t)\right).\notag
\end{align}
We construct a central finite difference/backward Euler time-stepper for (\ref{eq:moving})
as
\begin{equation}\label{eq:impl_euler}
U(t+\delta t) = S\left(U(t),\delta t\right) = \left[ I
- \delta t J\left(U(t)\right)\right]^{-1}U(t),
\end{equation}
where again $U(t)$ is the space discretization of $U(\zeta,t)$ on the lattice Boltzmann grid, and
\begin{displaymath}
J\left(U(t)\right) = \frac{\partial
\tilde{F}(U(t))}{\partial U}.
\end{displaymath}

We then define
\begin{equation}\label{eq:precond_mat}
M(\bar{U},\delta t) = \begin{bmatrix}
I - \left(I - \delta t J(\bar{U})\right)^{-1} & \d_{\zeta} \bar{U} \\
\d_{\bar{U}} p(\bar{U}) & 0
\end{bmatrix},
\end{equation}
with which we will solve linear systems of the form
\begin{equation}
M(\bar{U},\delta t)\begin{bmatrix} \d\bar{U} \\ \d
\alpha\end{bmatrix} = \begin{bmatrix} b_{\bar{U}}\\
b_{\alpha}\end{bmatrix}.
\end{equation}
The first $N$ equations can be simplified through some algebraic
manipulation,

\begin{displaymath}
\left[I-\left(I-\delta t J(\bar{U}\right)^{-1}\right] \d \bar{U}
+\d \alpha \cdot \d_{\zeta}\bar{U}=b_{\bar{U}},
\end{displaymath}
\begin{displaymath}
\d \bar{U} = \left(I-\delta t J(\bar{U}\right)^{-1} \d \bar{U} +
b_{\bar{U}} - \d \alpha \cdot \d_{\zeta}\bar{U},
\end{displaymath}
\begin{displaymath}
\d \bar{U} = \left(I-\delta t J(\bar{U}\right)^{-1} \d \bar{U} +
\tilde{b}_{\bar{U}},
\end{displaymath}
\begin{displaymath}
\left(I - \delta t J(\bar{U})\right)\d\bar{U} = \d \bar{U} +
\left(I - \delta t J(\bar{U})\right) \tilde{b}_{\bar{U}},
\end{displaymath}
\begin{displaymath}
\delta t J(\bar{U})\d\bar{U} = - \left(I - \delta t
J(\bar{U})\right)\tilde{b}_{\bar{U}},
\end{displaymath}
which leads to the equivalent linear system
\begin{equation}
\begin{bmatrix} \delta t J(\bar{U}) & -\left(I - \delta t J(\bar{U})\right)\d_{\zeta}\bar{U}\\
\d_{\bar{U}}p(\bar{U}) & 0
\end{bmatrix} \begin{bmatrix}
\d \bar{U}\\
\d \alpha
\end{bmatrix} = \begin{bmatrix}
- \left(I - \delta t J(\bar{U})\right)b_{\bar{U}}\\
b_{\alpha}
\end{bmatrix},
\end{equation}
in which the system matrix is the sum of a band and a rank one matrix.
There exist efficient (order $N$) direct solvers for linear systems of this type,
see e.g.~\cite{RafRef}.

\section{Numerical results\label{sec:6}}

\subsection{Fisher equation}

We consider the model problem \ref{mod:fisher} on the domain $[0,L]$ with $L=10$ and with
diffusion coefficient $a = 0.1$, and grid parameters $\d x = 2.5\cdot
10^{-2}$ and $\d t = 1\cdot 10^{-3}$.  For these parameters $N=400$
and $\omega \approx 1.35$. We intend to find the traveling wave with
minimal speed $c^*=2\sqrt{D}$, which is depicted in figure
\ref{fig:wave_fish}.  The initial guess $\bar{U}^{(0)}$ for the Newton
process is obtained by a time integration of the Fisher PDE
(\ref{eq:fish}) over a time interval of size $9000\d t$ with initial
condition
\begin{equation}
U(x,0)=\rho(x,0) = \exp(-5x^4), x \in [-L,L],
\end{equation}
from which we take the part where $x\ge 0$.

In order to choose an adequate value for the coarse-grained time-step $\delta t$,
we first illustrate the effects of the lifting operator.
We then investigate the spectrum of the linear system that we need to solve, along with
the effect of the preconditioner, and we conclude by showing the convergence of the Newton
process.
%
%
\subsubsection{Initialization}

To understand the details of the initialization, we compare the exact
evolution of the Fisher lattice Boltzmann model with the evolution
after re-initialization.  We perform an initial simulation of $100$
steps with the lattice Boltzmann model, starting from the initial
condition $\bar{U}^{(0)}$, which is lifted to distribution functions
using the first order slaving relations.  We extract the density at
$t=80\d t$ using equation (\ref{eq:lbm_mom}).  With this density, we
initialize a second lattice Boltzmann simulation using the three
procedures described in section \ref{sec:cg_ts}.  For the slaving
relations, we only use the first order approximation.  We then compare
the evolution of the distribution functions with the evolution of the
original simulation as we continue to evolve both. The result is
somewhat unexpected.
\begin{figure} \begin{center}
\includegraphics[scale=0.8]{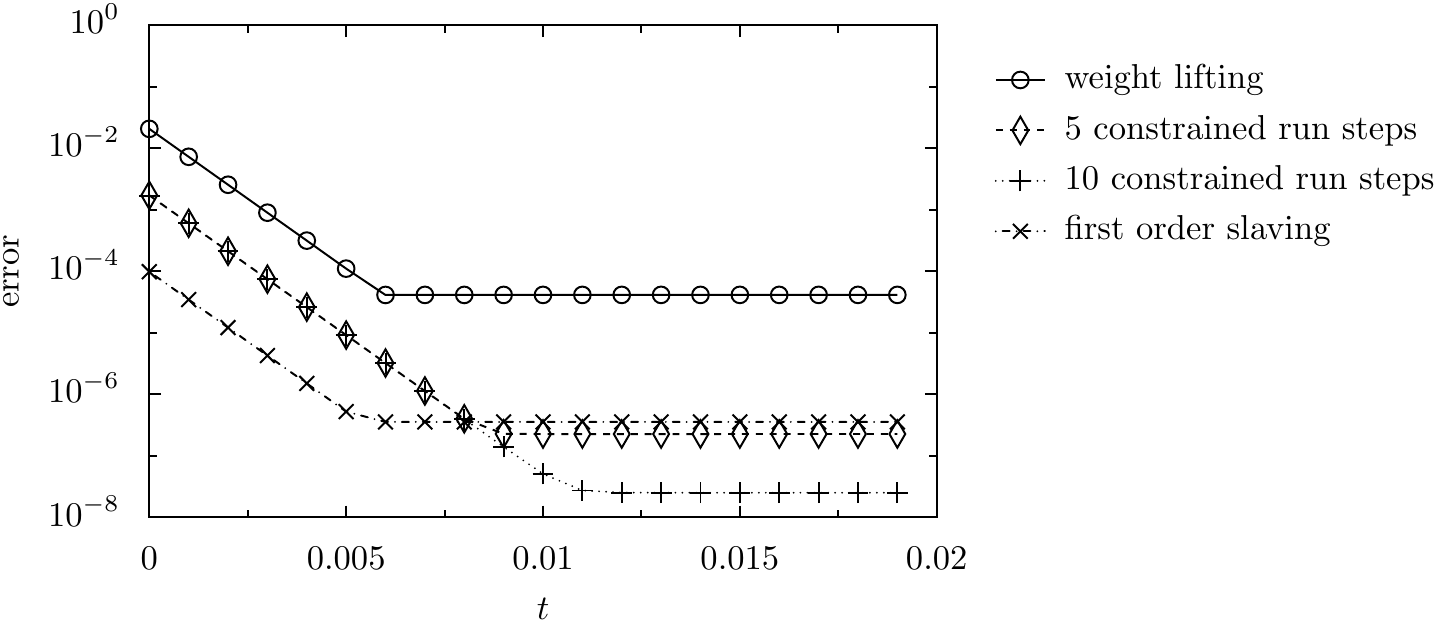}
\end{center}
\caption{\label{fig:fish_lift}
Evolution of the initialization error for the Fisher problem for different lifting procedures. We show the
difference between the distribution function of the re-initialized and
the original lattice Boltzmann simulation.
}
\end{figure}

In figure \ref{fig:fish_lift}, we plot the
error between the original solution and the re-initialized simulation in the
first $20$ time-steps after re-initialization.  We see that, in line with
the results reported in \cite{PvLWimRoo05}, both the constrained runs
scheme and initialization using first order slaving relations lead to
a reduced lifting error, compared to weighted lifting.
Unexpectedly, however, we observe that \emph{after} initialization, all
re-initialized simulations show an initial convergence towards the exact
solution with a convergence rate of $|1-\omega|$.  This convergence
stagnates after approximately 10 time-steps for this choice of $\omega$.
This behaviour is not completely understood yet.
As is to be expected, the smallest error is obtained by fully converged
constrained runs.

These results suggest that the simulation phase of the coarse-grained time-stepper,
which involves the
evolution of the lattice Boltzmann model from $t^*$ to $t^* + \delta t$ requires a
coarse-grained time-step $\delta t$ that is at least 10 times the microscopic time-step $\d t$ to eliminate initial
transients caused by the lifting for this choice of $\omega$.  If a shorter time is choosen,
undesired artifacts will show up in the spectrum of the coarse-grained
time stepper.  In our simulations, we choose $\delta t = 15\d t =
1.5\cdot 10^{-2}$, and we initialize using the constrained runs scheme
with $10$ steps.
%
%
\subsubsection{Performance of the preconditioner}\label{section:performance precond}
\begin{figure}
\begin{center} \includegraphics[scale=1]{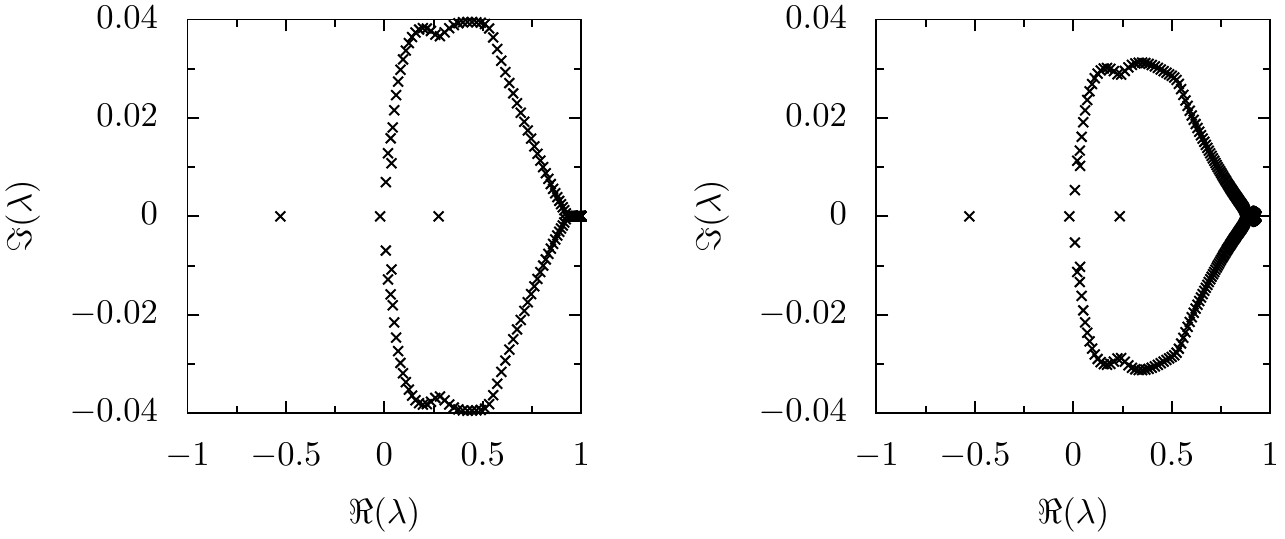} \end{center}
\caption{\label{fig:spec_fish}Fisher traveling wave problem.
Left: Spectrum of the system matrix of equation
(\ref{eq:linear}).  Right: spectrum of the preconditioning matrix (\ref{eq:precond_mat}).}
\end{figure}

We now compute the spectrum of the matrix $A(\bar{U}^{(0)},\delta t)$, using the
parameters defined above.    To this end,
we construct the matrix $A(\bar{U}^{(0)},\delta t)$ explicitly by computing the
matrix-vector products with all coordinate vectors $e_i = [0,\ldots, 0, 1, 0,
\ldots, 0]^T$. The matrix-vector products are estimated using (\ref{eq:matvec})
with $\epsilon=1\cdot 10^{-8}$.  The results are shown in figure
\ref{fig:spec_fish}(left).  We compare with the spectrum of the matrix
$M(U^{(0)},\delta t)$ shown in figure
\ref{fig:spec_fish}(right).  We see good agreement between the two spectra.
The differences are in the imaginary part of the eigenvalues, and in the very
fast modes, which show up around $1$ since we are computing the spectrum of the
fixed point iteration.  Also remark how the zero eigenvalue, corresponding to
the singularity of the fixed point equation (\ref{eq:sing}), 
is split into two isolated eigenvalues by the addition
of the phase condition and the artificial unknown $\alpha$.

\begin{figure}
\begin{center} \includegraphics[scale=1]{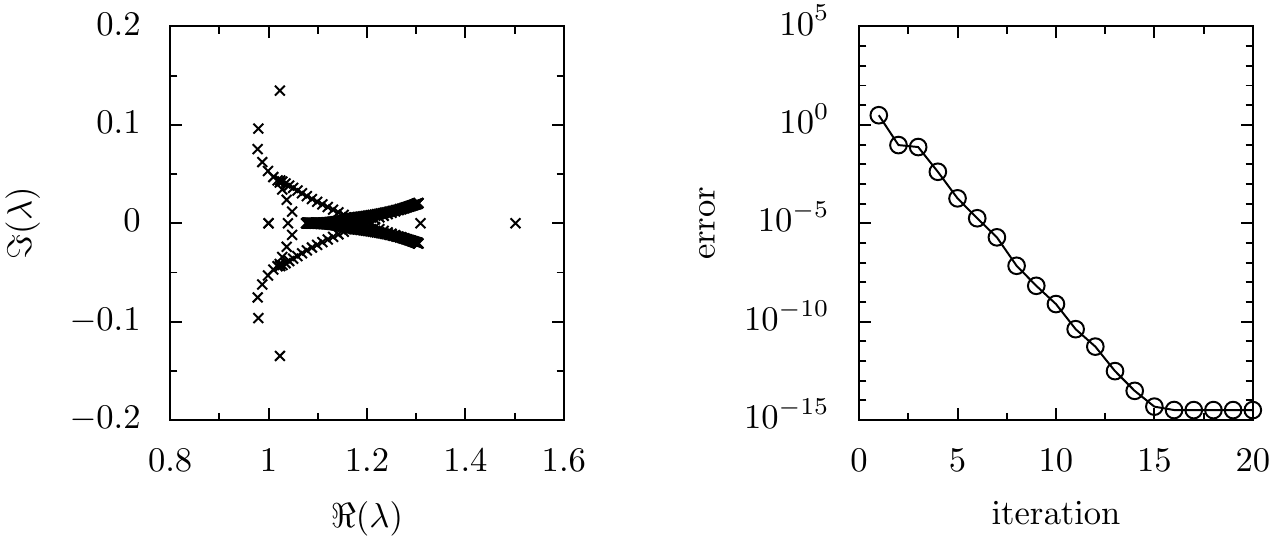} \end{center}
\caption{\label{fig:gmres_fish}Fisher traveling wave problem.
Left: spectrum of the product of the inverse of the preconditioner
and the system matrix.  Right: error of the linear system as a function of the number of GMRES iterations.}
\end{figure}

The spectrum of the matrix $M(\bar{U}^{(0)},\delta t)^{-1}A(\bar{U}^{(0)},\delta t)$ is shown in
figure \ref{fig:gmres_fish}(left).  We see that this spectrum is nicely clustered around $1$.
As a result, we obtain very fast GMRES convergence.  Figure \ref{fig:gmres_fish} shows the error
as a function of the number of GMRES iterations.  We clearly see the fast linear convergence,
as predicted by the theory \cite{saad}.
%
%
\subsubsection{Convergence of the Newton process}

\begin{figure}
\begin{center} \includegraphics[scale=1]{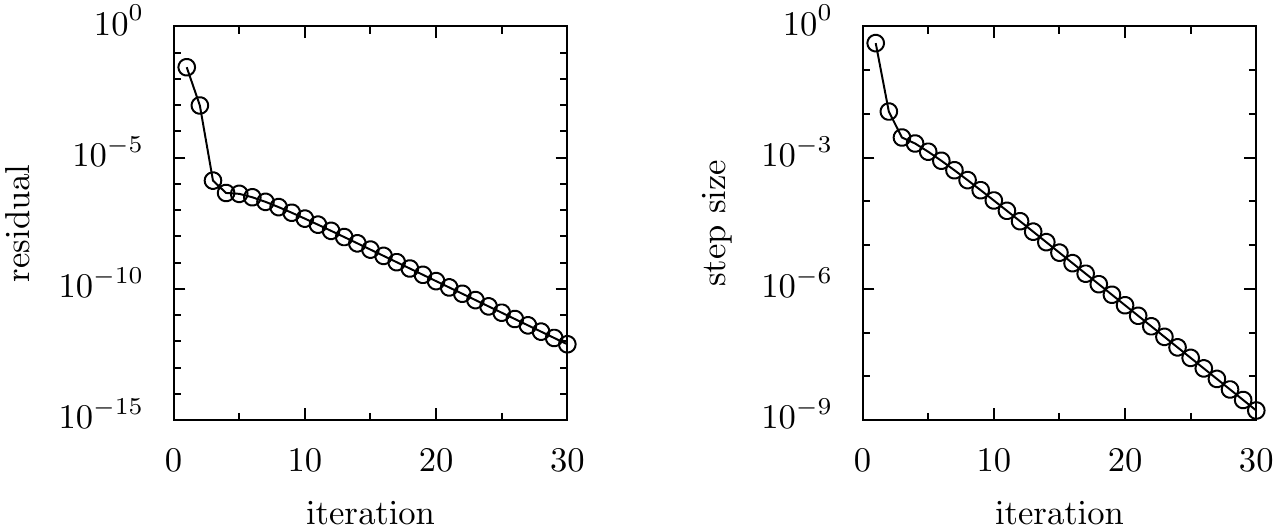}
\end{center}
\caption{\label{fig:newt_fish}Fisher traveling wave problem.
Convergence of the Newton process as a function of iteration number.
}
\end{figure}

We are now ready to show the convergence of the Newton process. To this end, we
take as an initial guess $\bar{U}^{(0)}$ a smoothed step function of the form
\begin{equation}
\bar{U}^{(0)} = \frac{1}{\exp(2(x-L/2))+1},
\end{equation}
with $L=10$, the length of the interval.
The convergence of the Newton process is shown in figure
\ref{fig:newt_fish}.  We notice that the convergence is linear, and
the convergence factor changes after approximately $4$ iterations.
In the process, we have converged to the traveling wave solution shown in figure
\ref{fig:wave_fish}, and $\alpha^*=-2.24\cdot 10^{-4}$.  The fact that $\alpha^*$ is
not equal to zero can be shown to be an artifact of the space discretization, combined
with the truncation to a finite domain.  Indeed, $\alpha^*$ becomes smaller
when we refine the lattice Boltzmann grid (and decrease the time-step to keep $\omega$ fixed).
We can explain this as follows: the conclusion that $\alpha^*$ should equal zero follows from the assumption
that the extra column that is added in the Jacobian, $\d_{\zeta}\bar{U}$, is the eigenvector
associated with the zero eigenvalue of the fixed point map (\ref{eq:sing}).  However, due to the
space discretization, $\d_{\zeta}\bar{U}$ will only be an \emph{approximation} to this
eigenvector.  Therefore, we are using a quasi-Newton method (hence the linear convergence), and
the solution value $\alpha^*$ will be non-zero but small.
The convergence factor changes when the traveling wave solution has been computed to
machine precision.  At that point, only the error in $\alpha$ shows up in
the residual, indicating that the non-zero value of $\alpha^*$ is indeed the
main cause for the non-quadratic convergence.

%
%
%

\subsection{Planar streamer front}
As a second example, we numerically study model problem \ref{mod:ionization} for planar ionization fronts
on the domain $[0,130]$ with grid parameter $\d x = 1\cdot 10^{-1}$ and $\d t = 1\cdot 10^{-3}$.
We choose the diffusion coefficient to be $a=1$ and the electric field $E^+_{\infty}=-1$ (i.e.~a constant, large
electric field on the far right).  For these parameters, the corresponding lattice Boltzmann
relaxation parameter is $\omega\approx1.538$, and $N = 1300$, which implies that the unknown solution has
$2601$ unknowns.  We intend to compute the traveling wave
solution with the minimal speed $c^*=|E^+_{\infty}|+2\sqrt{D g(E^+_{\infty})}$,
which is depicted in figure \ref{fig:wave_ebert}.
The initial guess $U^{(0)}$ for the Newton
process is obtained by a time integration of the ionization PDE (\ref{eq:ebert_pde})
over a time interval of size $4000\d t$ with initial condition
\begin{equation}
U(x,0)=\rho(x,0) = \exp(-15x^2), x \in [0,130].
\end{equation}

%
%
%
\subsubsection{Initialization}
\begin{figure}
\begin{center}
\includegraphics[scale=0.85]{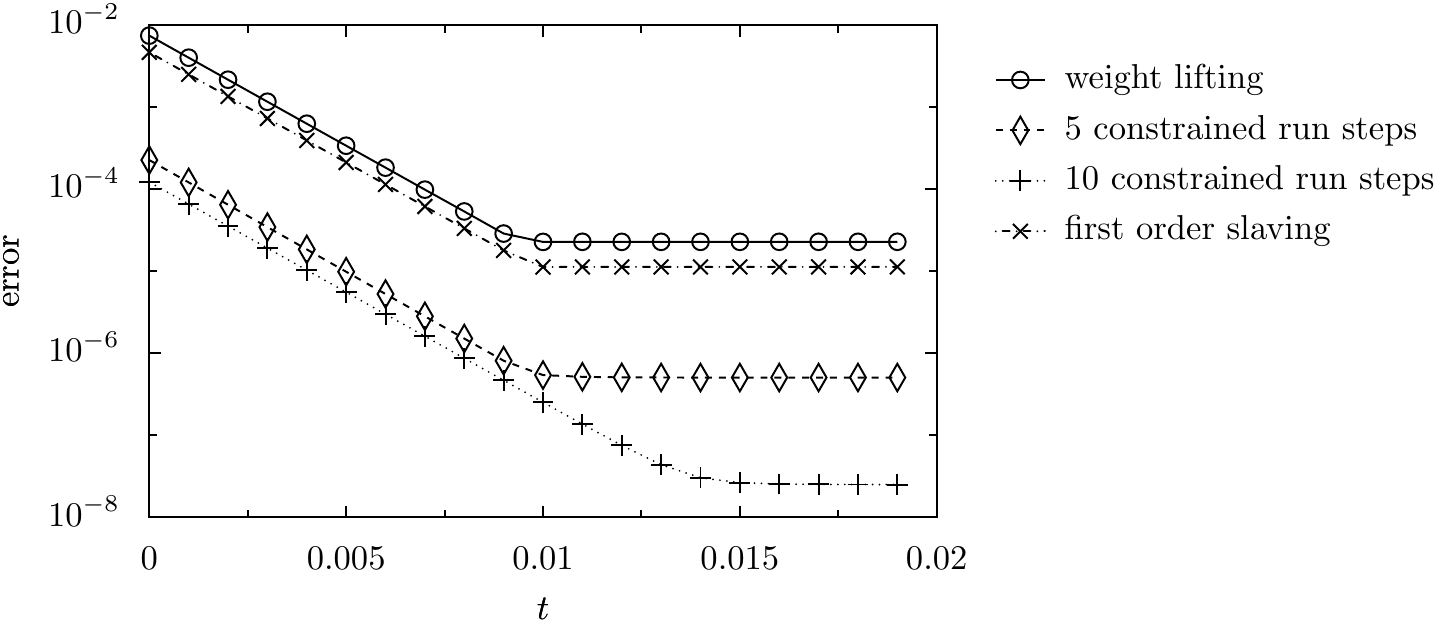}
\end{center}
\caption{\label{fig:ebert_lift}
Evolution of the initialization error for the planar streamer model with
different lifting procedures. We show the
difference between the distribution function of the re-initialized and
the original lattice Boltzmann simulation.}
\end{figure}

Again, we study the properties of the initialization by comparing the exact
evolution of a lattice Boltzmann simulation with the evolution after re-initialization.
We performed a simulation of $100$ steps, starting from $\bar{U}^{(0)}$, with
the lattice Boltzmann model (\ref{eq:lbm_ebert}), which is initialized using the
first order slaving relations.  Again, we extract the
density at time $t= 80\d t$, and initialize a second lattice Boltzmann simulation using
the three procedures described in section \ref{sec:cg_ts}.
We observe the same behaviour as for the Fisher model.
In figure \ref{fig:ebert_lift}, we show the evolution of the
initialization error during the first $20$ steps after re-initialization.  We again
see an initial convergence towards the exact solution, which stagnates
after $10$ to $20$ time-steps.  Depending on the accuracy of the lifting
step, we get a better correspondence with the correct distribution
functions.

Based on these observations, we choose $\delta t = 15\d t =
1.5\cdot 10^{-2}$ in our simulations, and we initialize using the constrained runs scheme
with $10$ steps.

\subsubsection{Performance of the preconditioner}
\begin{figure}
\begin{center} \includegraphics[scale=1]{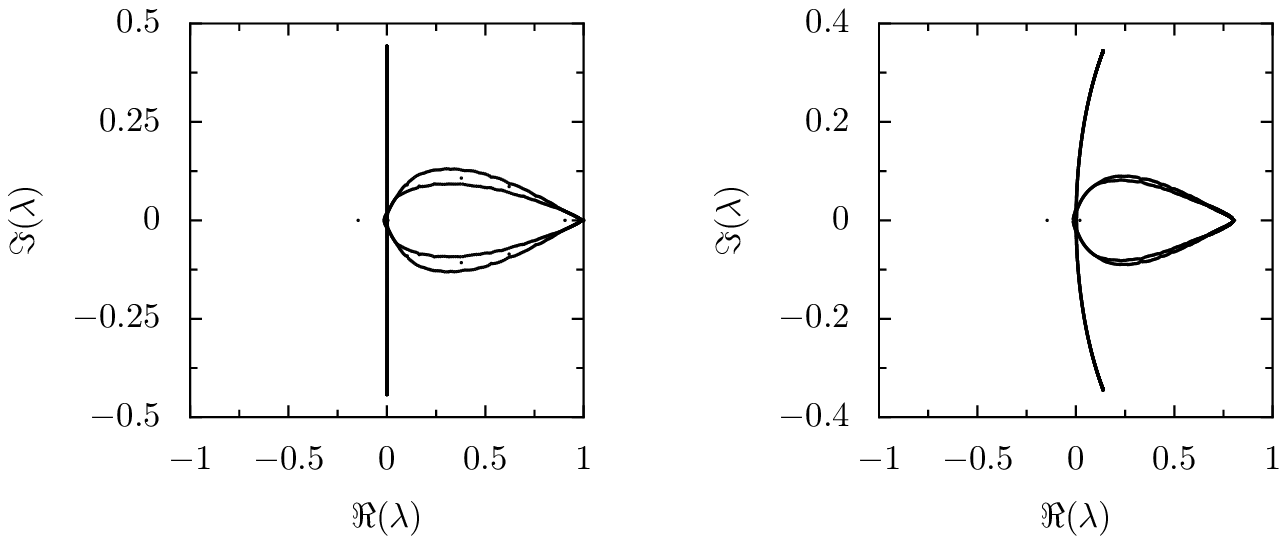}
 \end{center}
\caption{\label{fig:spec_ebert} The planar streamer front problem.
 Left: Spectrum of the
system matrix of equation (\ref{eq:linear}).  Right: spectrum of the
preconditioner (\ref{eq:precond_mat}).}
\end{figure}

In figure \ref{fig:spec_ebert}, we show the spectrum of the matrix
$A(\bar{U}^{(0)},\delta t)$, using the parameters defined above,
and the spectrum of the corresponding preconditioning matrix
$M(\bar{U}^{(0)},\delta t)$.
Again, we see good qualitative agreement between the two spectra, but the
differences are more pronounced than for the Fisher equation.

\begin{figure}
\begin{center}
\includegraphics[scale=1]{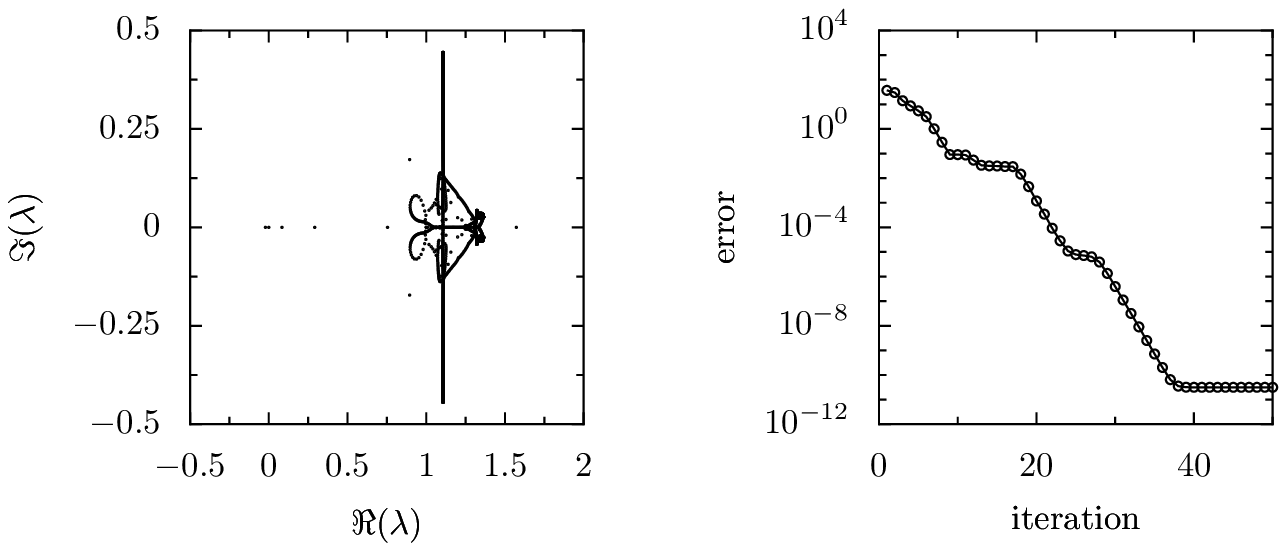}
 \end{center}
\caption{\label{fig:prec_ebert} Planar streamer front problem. Left: spectrum of the preconditioned
system matrix $M^{-1}A$ (equation (\ref{eq:precond})) for the first Newton step.
Right: the convergence of the preconditioned GMRES procedure as a function of the number of iterations.}
\end{figure}

The spectrum of the
matrix $M(\bar{U}^{(0)},\delta t)^{-1}A(\bar{U}^{(0)},\delta t)$ is shown in figure \ref{fig:prec_ebert}(left).  Although
most eigenvalues are in a small cluster around 1, a few eigenvalues
are scattered on the real axis.  These additional eigenvalues result in a
temporary stagnation of the GMRES convergence, as shown in figure
\ref{fig:prec_ebert} (right).  Between these stagnations,
linear convergence is observed.  We note that, even with the temporary stagnation,
the solution is computed in $30$ to $40$ iterations, while the number of unknowns is
$2601$.

\subsubsection{Convergence of the Newton process}
\begin{figure}
\begin{center}
\includegraphics[scale=1]{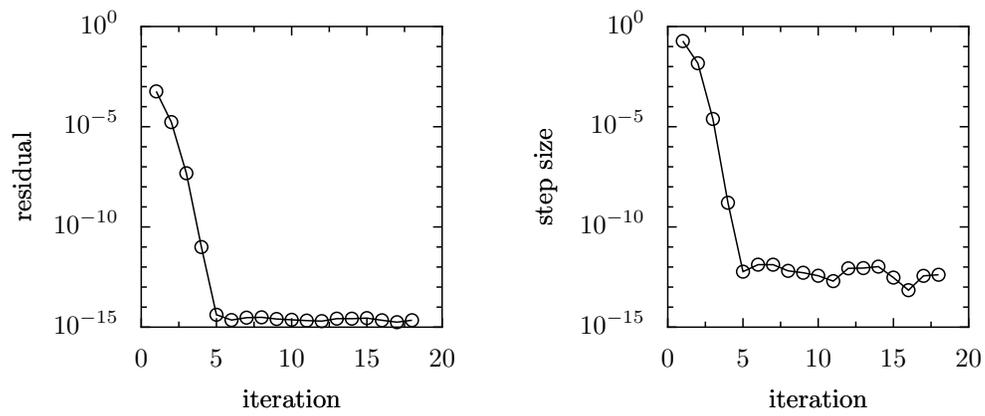}
\end{center}
\caption{\label{fig:newt_ebert}Planar streamer front problem.  Convergence of the Newton process
as a function of iteration number.}
\end{figure}

We proceed to show the convergence of the Newton process, starting from the initial
guess $\bar{U}^{(0)}$.  The results are shown in figure \ref{fig:newt_ebert}.
We plot the norm of the
residual and the correction $(\d \bar{U}^{(k)},\alpha^{(k)})$ after each iteration.
For this problem, we get quadratic convergence towards the solutions shown in figure
\ref{fig:wave_ebert}, and $\alpha^*=0$.  Thus, in this case, the effect of the discretization
is much less pronounced, or even absent.

%
%
%
%
\section{Discussion and conclusions\label{sec:7}}
In this article, we showed how one can use a coarse-grained
time-stepper to compute traveling wave solutions of lattice Boltzmann
models.  In a co-moving frame, emulated by performing a shift-back
operation after each coarse-grained time-step, the traveling wave
appears as a steady state, which is computed using a Jacobian-free
Newton--GMRES method.  The method uses repeated calls to the
coarse-grained time-stepper to estimate the required matrix-vector
products.  For efficiency reasons, we limited the size of the
coarse-grained time-step $\delta t$. The real part of the 
spectrum of the resulting Jacobian ranges from $0$ to $1$, which results
in slow convergence of Krylov methods. Therefore, we
accelerated convergence of the GMRES procedure by introducing a preconditioner that
is based on an approximate macroscopic model, which is derived from
the lattice Boltzmann model using a Chapman--Enskog expansion.  We
illustrated the approach on a lattice Boltzmann model for the Fisher
equation and on a model for an ionization wave.

The total cost of finding a solution is determined by the required number of
microscopic time-steps, which depends on the number of GMRES iterations and on
the number of microscopic time-steps per matrix-vector product. For each matrix-vector product, we use the
microscopic time-stepper both during the simulation and the lifting step when the constrained runs scheme
is used.  The required number
of calls to the microscopic time-stepper is determined by the microscopic relaxation parameter
$\omega$, since the convergence rate of the
distribution functions towards their correctly slaved values is given by $|1-\omega|$.  
In our examples, approximately $10$ lattice Boltzmann time-steps were needed for each
lifting operation and an additional $10$ to $15$ for each microscopic simulation.

The convergence rate of the GMRES method is closely related to the
quality of the preconditioner.  The preconditioner performs well when
the approximate macroscopic model is a good approximation of the
lattice Boltzmann model.  Our experiments indicate that the quality of the preconditioner
is not related to the number of lattice points $N$, which implies that the number of
GMRES iterations is approximately constant with varying $N$. 

Since the number of time-steps only depends on the
relaxation parameter $\omega$ and the spectrum of the preconditioned
time-stepper, we conclude that the convergence rate is independent of the number of
variables in the problem.  As a consequence, the algorithm scales with
the cost of taking a single lattice Boltzmann time-step and 
the cost of a direct solve with the preconditioning matrix.

In our numerical examples we found that, on average, 2000 time-steps
are required to converge to the traveling wave solution. This is
still expensive if we compare with straightforward time
integration.  E.g.~for the Fisher model, a localized initial state
evolves into a steady traveling wave within 9000 time-steps.
However, direct time integration looses its appeal in situations where
the microscopic time-step is so small that it is computationally
infeasible to reach the time horizon where the traveling wave becomes
steady.  Furthermore, the total number of time-steps will be much
smaller in the context of continuation, where the behaviour of the
solutions is studied as a function of one or more varying
parameters. In this setting, the solution for nearby parameter values
provides an accurate initial guess, hereby lowering the required
number of Newton steps.  Also, the Newton--GMRES method allows us to
find \emph{unstable} traveling waves.  These will never be found by
simulation, since any perturbation will grow, and will destroy the
traveling wave.

Note that preconditioning with a roughly approximate macroscopic PDE model can
also be helpful to study the linear stability properties of the
traveling wave.   Indeed, the relevant rightmost eigenvalues of the Jacobian
matrix in the solution can be computed by an Arnoldi
iteration that requires the solution of a linear system in each iteration
\cite{LeHSal01}. This linear system can again be preconditioned with the
techniques proposed in this article.

In principle, the Newton--GMRES method could be applied to the lattice
Boltzmann model directly.  However, it is much harder to find an
appropriate preconditioner in that setting.  Indeed, one cannot expect
a preconditioner based on the approximate PDE to work well, since the
fast time-scales, which describe the slaving of the higher order
moments, will not be correctly accounted for.  We also emphasize that,
from the coarse-grained solution, the detailed solution can easily be
reconstructed using the constrained runs lifting scheme.  Our method
generalizes readily to lattice Boltzmann models with a more detailed
description of the velocity space.  A particular five-speed model, in
which the Townsend approximation for the reaction term is replaced by
a more realistic set of microscopic interaction rules, is currently
being investigated \cite{WimSamPvLRoo06}.

\section*{Acknowledgements}
The authors thank Pieter Van Leemput, Christophe Vandekerckhove, Kurt
Lust and Tim Boonen for interesting discussions about various aspects
of this work.  Pieter Van Leemput provided us with figure
\ref{fig:constrained}.  
GS is a Research Assistant of the Fund for Scientific
Research -- Flanders.  WV is supported by the Belgian Science Policy Office
through its action ``return grants''.  
This work was supported by the Fund for Scientific Research -- Flanders
through Research Project G.0130.03 (GS, DR, WV) and by the Interuniversity 
Attraction Poles Programme of the Belgian Science Policy Office through grant 
IUAP/V/22. The scientific responsibility rests with its authors.
The research of IGK was partially supported by the US DOE and DARPA.  

\bibliographystyle{plain}

\end{document}